\documentclass[preprint,floats,twoside,eqsecnum,aps]{revtex4}
\usepackage{graphicx}
\begin{document}


  \title{ DERIVATIVE DISPERSION RELATIONS FOR THE AMPLITUDE SLOPES
   IN  pp AND $\bar{\rm p}$p SCATTERING }

\author{ ERASMO FERREIRA }
\address{Universidade Federal do Rio de Janeiro,  P.O.Box 68528
Rio de Janeiro, RJ 22461-200, Brazil  \\
erasmo@if.ufrj.br}

\begin{abstract}
 We extend the use of derivative dispersion relations to the study 
of slopes of the real and imaginary amplitudes in 
pp  and $\bar{\rm p}$p elastic scattering. The new relations are 
tested against the solutions for the amplitudes obtained in the 
analysis of the high energy data. Extensions beyond the forward 
direction are investigated.
 \end{abstract}
\maketitle
  
  \section{Introduction}
Derivative Dispersion Relations (DDR) for $|t|=0$ have received much 
interest \cite{menon,nicolescu}, but their extensions to
general momentum transfers  $|t|$  have  been 
very limited \cite{Bronzan}, due, in the practical aspects, to the 
absence of knowledge on the imaginary and real hadronic amplitudes 
for arbitrary $t$. However, a treatment of the 
 $\rm pp$ and ${\bar{\rm p}{\rm p}} $ data  
has proposed a solution for the   amplitudes \cite{flavio1,flavio3}, 
which is likely realistic as a phenomenological description. 
  This solution can be tested through DDR written for arbitrary 
$|t|$, and this is the purpose of the present work. 

 The momentum transfer $t$ enters as a parameter in the $s$-channel 
dispersion relations. Thus in a simplest form of DDR, to be applied
to pp and $\bar {\rm p}$p scattering at high energies, 
 we write \cite{menon,nicolescu},  
for the even and odd  combinations  
\begin{equation}
\label{ddrplus}
\frac{{\rm Re} F_+(s,t)}{s}= \frac{K}{s} + 
\Bigg[\tan \bigg(\frac{\pi}{2} \frac{d}{d\log{s}}\bigg ) 
    \Bigg] \bigg[\frac{{\rm Im} F_+(s,t)}{s}\bigg],  
\end{equation}
 \begin{equation}
\label{ddrminus2}
\frac{\pi}{2}\frac{d}{d\log{s}}\bigg[\frac{{\rm Re} F_-(s,t)}{s}\bigg]
=-\Bigg[ \bigg(\frac{\pi}{2} \frac{d}{d\log{s}}\bigg) 
\cot \bigg(\frac{\pi}{2} \frac{d}{d\log{s}}\bigg) \Bigg]
\bigg[ \frac{{\rm Im} F_-(s,t)}{s}\bigg] ~ .
\end{equation}
 The amplitudes for the $pp$ and $\bar p p$ channels  are  
 $F_{\rm pp}=F_++F_- ~ ~ ~ {\rm and} ~ ~ ~ F_{\bar{\rm p}{\rm p}}=F_+-F_- ~ .$   
  The normalization  for each channel 
(pp or $\bar{\rm p}$p) is defined by the optical 
theorem $ \sigma(s)=  {{\rm Im} F(s,t=0)}/{s}  ~.$
The ratio $ \rho (s) ={\rm Re}F(s,t=0)/{{\rm Im}F(s,t=0)}$ 
of real to imaginary amplitude at $t=0$ 
and the exponential slope  
are characteristic parameters of elastic scattering. With amplitudes  
described 
at low $|t|$ by exponential slopes, we write   for each channel, 
$ {\rm Re} F(s,t)={\rm Re} F(s,0) \exp (-{B^R} |t|/2)$, 
$ {\rm Im} F(s,t)={\rm Im} F(s,0) \exp (-{B^I} |t|/2) ~ . $
 \section{ DDR for $ t=0$}
In terms of experimentally measured quantities the DDR for the 
even and odd amplitudes with $ t=0 $  are written respectively
\begin{equation}
\label{ddreven}
\frac{1}{2}\bigg( \sigma_{\rm pp} \rho_{\rm pp}
     +\sigma_{\bar{\rm p}{\rm p}}\rho_{\bar{\rm p}{\rm p}} \bigg)
   =\frac{K}{s}+\frac{1}{2}\Bigg[
\frac{\pi}{2}\frac{d}{d\log{s}}
 +\frac{1}{3}\bigg(\frac{\pi}{2}\frac{d}{d\log{s}}\bigg)^3
  + \dots \Bigg]
 \bigg[\sigma_{\rm pp}+\sigma_{\bar{\rm p}{\rm p}}\bigg] ~ , 
\end{equation}
\begin{equation}
\label{ddrodd}
\frac{1}{2}\frac{\pi}{2}\frac{d}{d\log{s}}\bigg[\sigma_{\rm pp}\rho_{\rm pp}
-\sigma_{\bar{\rm p}{\rm p}}\rho_{\bar{\rm p}{\rm p}}\bigg] = 
-\frac{1}{2} \Bigg[
1-\frac{1}{3}\bigg(\frac{\pi}{2} \frac{d}{d\log{s}}\bigg )^2
 - \dots \Bigg]\bigg[ 
\sigma_{\rm pp}-\sigma_{\bar{\rm p}{\rm p}}\bigg] ~ . 
\end{equation}

  We parametrize the total   cross-sections with two alternative forms 
\begin{equation}
\label{cross1}
\sigma=  D+d_0 \log^2(s/s_0) +  d_1 s^{-\mu_1}- \tau a_2 s^{-\eta_2}~ , 
\end{equation}
\begin{equation}
\label{cross2}
{\rm and} ~ ~ ~ ~ ~ \sigma= a_0  s^\epsilon +  a_1 s^{-\eta_1}- \tau  a_2 s^{-\eta_2} ~ .
\end{equation} 
where $\tau=$ +1 and -1  for pp and 
$\bar{\rm p}$p respectively, $s$ is in GeV$^2$, 
and $s_0=25 ~ {\rm GeV}^2$.
 
Using these representations for $\sigma$, we use  
$  d^n s^\lambda/d\log{s}^n=\lambda ^n s^\lambda $,  and 
   may write in closed forms the RHS terms of Eqs. (\ref{ddreven}), 
(\ref{ddrodd}). We thus have for the even-DDR 
\begin{equation}
\label{ddreven2c}
\frac{1}{2}\bigg( \sigma_{\rm pp} \rho_{\rm pp}
     +\sigma_{\bar{\rm p}{\rm p}}\rho_{\bar{\rm p}{\rm p}} \bigg)
 =\frac{K}{s}+2 d_0 \bigg(\frac{\pi}{2}\bigg)\log(s/s_0)
 -d_1 \tan{\bigg(\frac{\pi}{2} \mu_1\bigg)}s^{-\mu_1} ~ , 
 \end{equation}
\begin{equation}
\label{ddreven2a}
\frac{1}{2}\bigg( \sigma_{\rm pp} \rho_{\rm pp}
     +\sigma_{\bar{\rm p}{\rm p}}\rho_{\bar{\rm p}{\rm p}} \bigg)
   =\frac{K}{s}+ a_0 \tan{\bigg(\frac{\pi}{2}\epsilon\bigg)}s^{\epsilon}
    - a_1 \tan{\bigg(\frac{\pi}{2} \eta_1\bigg)}s^{-\eta_1} ~ , 
\end{equation}
respectively for the  squared log and power representations of the 
cross sections. The series would not converge for 
non-integer powers of $\log{s}$. 

 The RHS of the odd-DDR is the same for both representations,
\begin{equation}
\label{ddrodd2}
{\rm RHS(odd)}= a_2 \Bigg[
1-\frac{1}{3}\bigg(\frac{\pi}{2} \eta_2 \bigg)^2
 - \dots \Bigg] s^{-\eta_2} 
= a_2  \bigg(\frac{\pi}{2}\eta_2 \bigg) 
 \cot{\bigg(\frac{\pi}{2}\eta_2 \bigg)} s^{-\eta_2} ~  . 
\end{equation}

Then  the whole odd-DDR in Eq.(\ref{ddrodd}) can be integrated
(from infinity to $s$), giving 
\begin{eqnarray}
\label{oddintegrated}
\frac{1}{2}\bigg[\sigma_{\rm pp}\rho_{\rm pp}
-\sigma_{\bar{\rm p}{\rm p}}\rho_{\bar{\rm p}{\rm p}}\bigg] =
- a_2 \cot{\bigg(\frac{\pi}{2}\eta_2 \bigg)} 
s^{-\eta_2}    ~ 
= \frac{1}{2} \cot{\bigg(\frac{\pi}{2}\eta_2 \bigg)} 
\bigg[\sigma_{\rm pp}-\sigma_{\bar{\rm p}{\rm p}}\bigg] ~ .  
\end{eqnarray}
This  relation connects  observable quantities, without derivative 
operations. These 
DDR for $|t|=0$ have been  tested 
against experimental values of  $\rho_{\rm pp}$ and 
$\rho_{\bar{\rm p}{\rm p}}$.
\section{ DDR for amplitude slopes }    
The purpose of the present work is to examine the   
validity of DDR beyond the forward direction. We here 
present results for the amplitude slopes.
Taking the exponential forms for the amplitudes  
into Eq.(\ref{ddrplus}), expanding for small $|t|$, and 
taking into account that DDR  at $|t|=0$ are valid , we obtain
from the even DDR, to first order in $|t|$ ,  
\begin{equation}
\label{slopeeven}
   B^R_{\rm pp} \sigma_{\rm pp} \rho_{\rm pp} +
  B^R_{\bar{\rm p}{\rm p}} \sigma_{\bar{\rm p}{\rm p}}\rho_{\bar{\rm p}{\rm p}}  
 = \tan \bigg[ \frac{\pi}{2} \frac{d}{d\log {s}} \bigg]  \big[
B^I_{\rm pp} \sigma_{\rm pp} +
   B^I_{\bar{\rm p}{\rm p}} \sigma_{\bar{\rm p}{\rm p}} \big] ~ .
\end{equation}
Analogously   from the odd DDR we obtain
\begin{equation}
\label{slopeodd}
 \frac {d}{d\log {s}}\bigg[ B^R_{\rm pp}\rho_{\rm pp}\sigma_{\rm pp} 
-B^R_{\bar{\rm p}{\rm p}} \rho_{\bar{\rm p}{\rm p}} \sigma_{\bar{\rm p}{\rm p}}\bigg]=
-\bigg[ (\frac{d}{d\log {s}}) \cot(\frac{\pi}{2} \frac{d}{d\log {s}})\bigg]
\big[ B^I_{\rm pp} \sigma_{\rm pp} 
- B^I_{\bar{\rm p}{\rm p}}\sigma_{\bar{\rm p}{\rm p}} \big] ~. 
\end{equation}

To evaluate these expressions we need to know the amplitude slopes and 
their energy dependences. Following the results obtained in the analysis of 
data \cite{flavio3}, we assume that the imaginary slopes  are  the same for 
the pp and ${\bar{\rm p}{\rm p}}$ channels, with linear dependences in 
$\log {s}$. Thus
\begin{eqnarray}
\label{slopes}
 B^I_{\rm pp}=B^I_{\bar{\rm p}{\rm p}}=B^I(s)=c_1^I+c_2^I~\log {s} ~ . 
\end{eqnarray}
 Using the relations
$$ \frac{d^n}{d(a x)^n} \big[x Z(x)\big]=
    \frac{n}{a} \frac{d^{n-1}}{d(a x)^{n-1}} \big[Z(x)\big ]+
       x \frac{d^n}{d(a x)^n }\big[ Z(x)\big]  $$
we can operate with the series expansions and obtain closed 
forms for the slope DDR.  The expressions that we need are 
$$      \tan    ( \frac{\pi}{2}\frac{d}{d \log{s}} )\big[ \log{s} ~  Z(s)\big]=
\log{s} ~  \tan ( \frac{\pi}{2}\frac{d}{d \log{s}} )\big[ Z(s)\big] +
\frac{\pi}{2} \sec ^2 ( \frac{\pi}{2}\frac{d}{d \log{s}} )\big[ Z(s) \big]  ~ , $$
and 
$$ \frac{\pi}{2}\frac{d}{d \log{s}} \cot ( \frac{\pi}{2}\frac{d}{d \log{s}} )
\big[ \log{s} ~  Z(s)\big] =  $$
 $$  \log{s} ~
\frac{\pi}{2}\frac{d}{d \log{s}} \cot ( \frac{\pi}{2}\frac{d}{d \log{s}} )\big[ Z(s)\big] +
\frac{\pi}{2}\big[ \cot (\frac{\pi}{2}\frac{d}{d \log{s}})-
\frac{\pi}{2}\frac{d}{d \log{s}}
 {\rm cosec} ^2 ( \frac{\pi}{2}\frac{d}{d \log{s}} ) \big] \big[ Z(s) \big]  ~ . $$

To study the odd DDR, we use the 
parametrizations of $\sigma$,  and   write  
 \begin{eqnarray}
\label{slopeodd2}
 \frac{\pi}{2} \frac {d}{d\log {s}}\bigg[ B^R_{\rm pp}\rho_{\rm pp}\sigma_{\rm pp} 
-B^R_{\bar{\rm p}{\rm p}} \rho_{\bar{\rm p}{\rm p}} \sigma_{\bar{\rm p}{\rm p}}\bigg]
=    \nonumber \\  
-  (\frac{\pi}{2}\frac{d}{d\log {s}}) \cot(\frac{\pi}{2} \frac{d}{d\log {s}}) 
\bigg[ (c_1^I  + c_2^I \log (s)) (\sigma_{\rm pp}- \sigma_{\bar{\rm p}{\rm p}} ) \bigg]= 
     \nonumber \\
 2 a_2 ~ s^{-\eta_2}\bigg[ B^I \big[\frac{\pi}{2}\eta_2 ~  \cot (\frac{\pi}{2}\eta_2 )\big] 
+ c_2^I ~ \frac{\pi}{2}\big[-\cot (\frac{\pi}{2}\eta_2)+\frac{\pi}{2}\eta_2 ~ 
{\rm cosec} ^2(\frac{\pi}{2}\eta_2)\big] \bigg] ~. 
\end{eqnarray}
    Integrating both sides from infinity to $s$ we have for the odd DDR for slopes  
 \begin{equation}
\label{slopeodd3}
  B^R_{\rm pp}\rho_{\rm pp}\sigma_{\rm pp} 
-B^R_{\bar{\rm p}{\rm p}} \rho_{\bar{\rm p}{\rm p}} \sigma_{\bar{\rm p}{\rm p}}=   
-2 a_2  ~ s^{-\eta_2}\big[ B^I \cot(\frac{\pi}{2}\eta_2)
+ c_2^I ~ \frac{\pi}{2}~  {\rm cosec^2}(\frac{\pi}{2}\eta_2)\big] ~ .
\end{equation}
  
 For the even DDR for slopes  we have to consider two cases separately.

I- ~  $~ \sigma$ of $ \log ^2(s) $ kind, Eq.(\ref{cross1}). We obtain
\begin{eqnarray}
\label{evenslopeI}    
    B^R_{\rm pp} \sigma_{\rm pp} \rho_{\rm pp} +
  B^R_{\bar{\rm p}{\rm p}} \sigma_{\bar{\rm p}{\rm p}}\rho_{\bar{\rm p}{\rm p}}=  
B^I \big[ 2\pi d_0 \log (s/s_0)- 2 d_1 \tan (\frac{\pi}{2} \mu_1) s^{-\mu_1}\big]
                 \nonumber \\
+ c_2^I ~ \pi ~ \big[  d_0  \log ^2(s/s_0)+ d_0  \frac{\pi^2}{2} + D +
 d_1 \sec ^2 (\frac{\pi}{2} \mu_1) s^{-\mu_1} \big] ~ .
\end{eqnarray}
 II-~  $~ \sigma$ of $ s^\epsilon $ kind, Eq. (\ref{cross2}). We here obtain
\begin{eqnarray}
\label{evenslopeII}    
 B^R_{\rm pp} \sigma_{\rm pp} \rho_{\rm pp} +
 B^R_{\bar{\rm p}{\rm p}} \sigma_{\bar{\rm p}{\rm p}}\rho_{\bar{\rm p}{\rm p}}=  
 B^I \big[ 2 a_0 \tan (\frac{\pi}{2}\epsilon)~ s^\epsilon 
- 2 a_1 \tan(\frac{\pi}{2} \eta_1)~ s^{-\eta_1}\big] \nonumber \\
+ c_2^I ~ \pi ~ \big[  a_0 \sec^2 (\frac{\pi}{2}\epsilon) ~ s^\epsilon
+  a_1 \sec ^2 (\frac{\pi}{2}\eta_1)~ s^{-\eta_1} \big] ~ .
\end{eqnarray}

\begin{figure}[ht]
 \vskip 2mm
 \includegraphics[height=6cm,width=6cm] {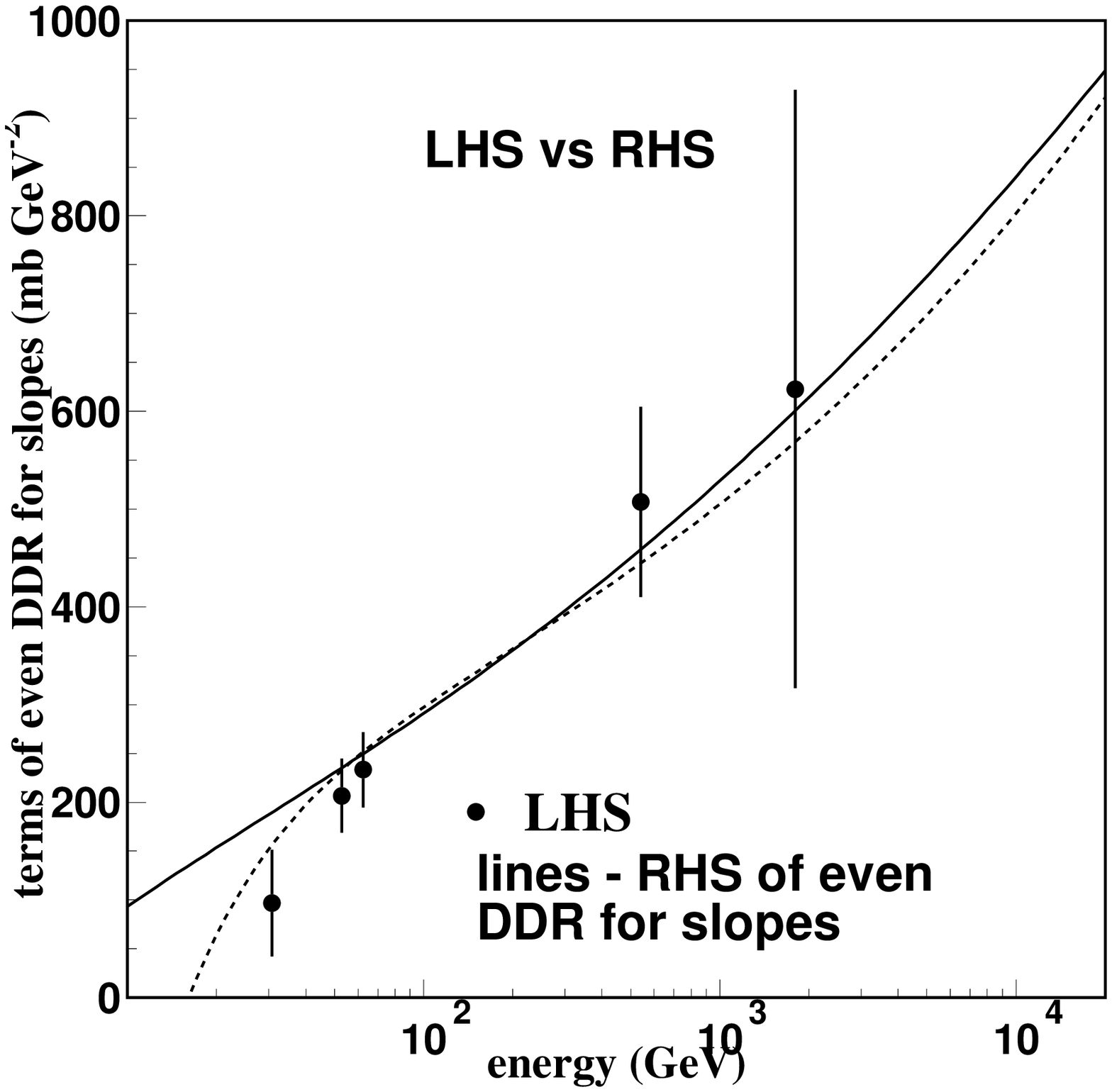 }
 \includegraphics[height=6cm,width=6cm]{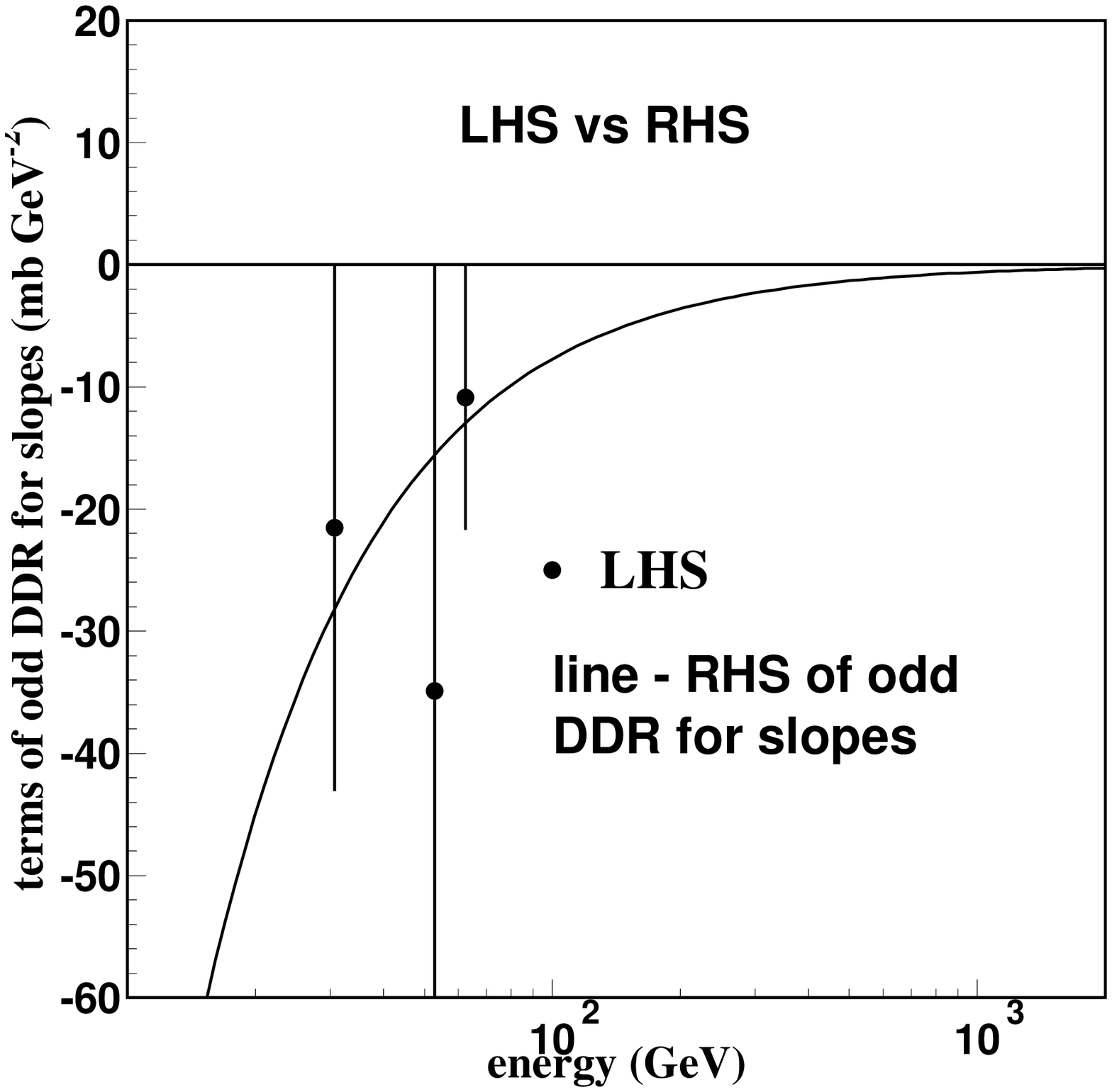 }
   \label{ddrslopes_fig}
    \caption{  
  The experimental points  (full circles end error bars)represent 
the LHS of the even and odd DDR for slopes. The lines represent the RHS 
of Eqs. (\ref{evenslopeI}), 
(\ref{evenslopeII})  and (\ref{slopeodd3}) , with (for the even case) 
full line for $\sigma$  of form $\log^2(s/s_0)$ and   dashed  line 
for the form with $s^\epsilon$. } 
\end{figure}

 Knowing the imaginary slope we may draw lines representing the RHS of 
the DDR for slopes. The LHS 
 is given by the data, using $B^R$ slopes of the pp  and 
$ {\bar{\rm p}{\rm p}}$ amplitudes \cite{flavio3}.
 The results are shown in  Fig. \ref{ddrslopes_fig}, where we observe 
compatibility, confirming both the values of the slopes and
the adequacy of the new dispersion relations.
  \section*{Acknowledgements}
The author is grateful to CNPq (Brazil) for support of his 
scientific research.

\end{document}